\documentclass[aps,prl,reprint]{revtex4-2}
\usepackage{mathtools}
\usepackage{graphicx}
\usepackage{dblfloatfix}
\usepackage{upgreek}
\begin{document}

\title{Polarisation Control of Nanoantenna Cold Spots}
\author{Alexander J. Vernon}
\email{alexander.vernon@kcl.ac.uk}
\author{Francisco J. Rodr\'iguez-Fortu\~no}
\email{francisco.rodriguez-fortuno@kcl.ac.uk}
\affiliation{Department of Physics, King's College London, Strand, London WC2R 2LS, United Kingdom}

\date{\today}

\begin{abstract}
Cold spots are sub-wavelength regions which might emerge near a plasmonic nanoantenna, should one or more components of some far-field illumination cancel out with scattered light. With a simplest-case demonstration using two dipolar scatterers, we show that by changing only the polarisation and amplitude of two plane waves, a unique, zero-magnitude and super sub-wavelength cold spot can be created anywhere in the space around a nanoantenna. This technique is a means for ultra-fast, remote, and non-mechanical sub-wavelength electric field manipulation.
\end{abstract}
\maketitle
If visible light shines on a very small metal particle, power will be scattered in different directions. By changing the particle’s shape, much of this power can be squeezed into confined pockets of space, referred to as hot spots \cite{novotny06, maier06}. In exchange, dimmer regions develop around the particle where the scattered and incident light begin to cancel out, and any dark zone where one or more components of interest are completely suppressed is called a cold spot \cite{haggui12, zito16}.

Plasmonic nanoantennas are engineered to scatter optical-frequency light in a valuable way \cite{giannini11, bharadwaj09}. As research interest has grown rapidly in the recent decades, many structures have been designed which brighten received light in hot spots by a factor of over 1000. Such highly intense hot spots have been applied to photovoltaics \cite{schuller10, atwater10, tan12, paetzold15, abdelraouf18, mohsen20}, and are a natural asset to fluorescence \cite{li17, kinkhabwala09} and surface-enhanced Raman scattering spectroscopy \cite{camden08, li10, jackson04, nie97} as well as biosensing and other nanomedicine applications \cite{lee08, anker08, law11, mayer11, wijaya09}. Meanwhile, with a highly suppressed electric field in contrast to a hot spot, cold spots have been used to pinpoint single fluorescent molecules in MINFLUX microscopy \cite{balzarotti17}. A molecule can be located by counting the number of photons it emits as it is probed for by a doughnut beam with a cold spot at its centre, because the beam’s shape and position are known (should the molecule lie in the cold spot, it will not release any photons). If a hot spot can be thought of as an ‘on’ switch for a quantum dot emitter, then a cold spot is its ‘off’ switch. Changing the handedness of some elliptically polarised incident light in \cite{tang18}, the field in some regions near a nanostructure was either strongly enhanced or suppressed so that two quantum dots could be selectively addressed. An alternative application for cold spots might exploit the fact that a beam, blue-detuned with respect to a certain electron transition, can push atoms into wells of low electric field intensity \cite{vetsch10}. If an illuminated plasmonic nanostructure unlocks the ability to manipulate an electric field of appropriate wavelength, atoms could be trapped and easily manoeuvred in controllable cold spots.
\begin{figure}
\includegraphics[width=0.5\textwidth]{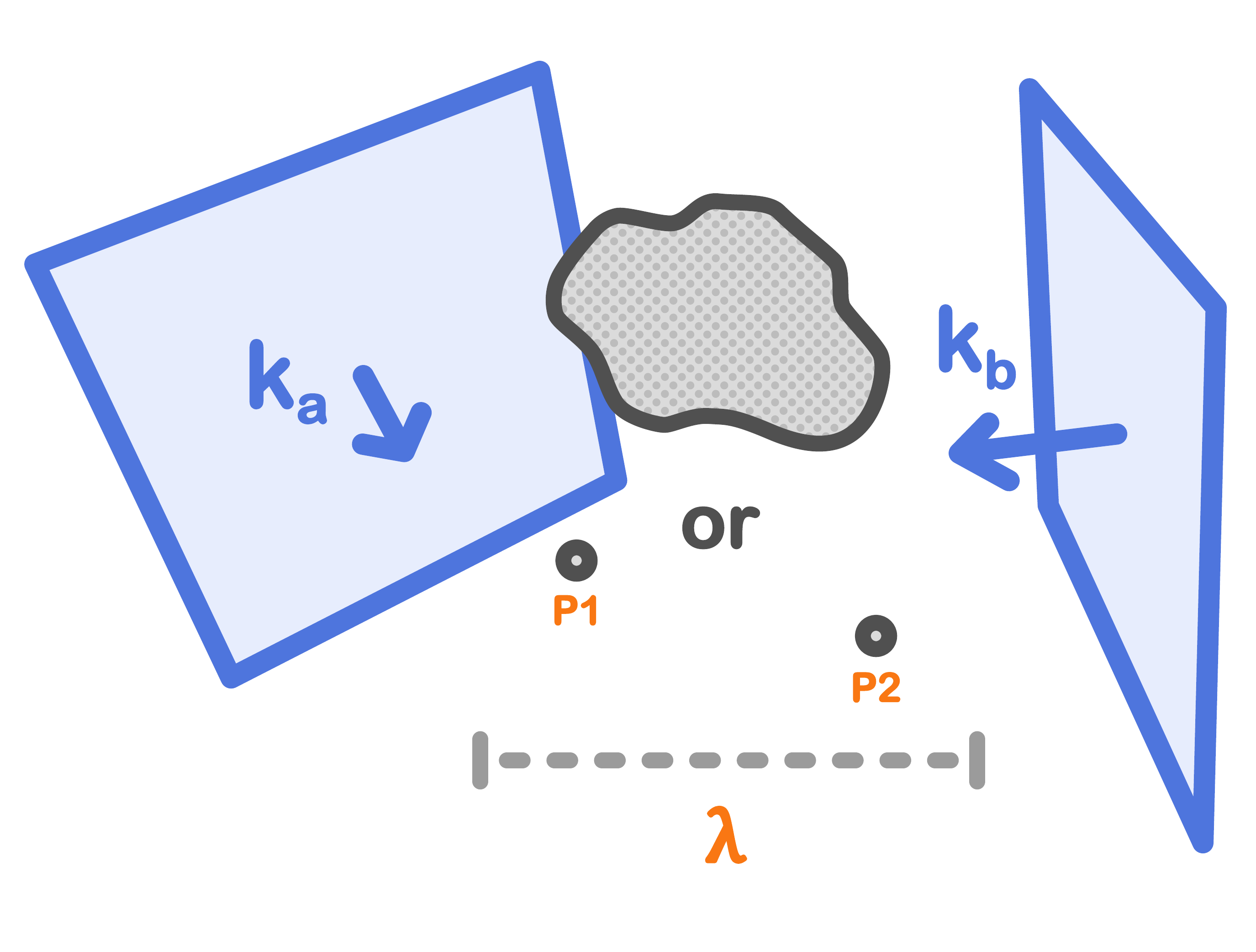}
\caption{
Illustration of two plane waves incident on a larger scatterer (comparable in size to $\lambda$) or two separated point particles P1 and P2.
}
\end{figure}

Whether or not a hot or cold spot emerges in a nanoantenna’s near field depends on its geometry, its material properties, and the characteristics of the exciting light. The phase shift introduced by the nanoantenna as it scatters light is tuneable by changing the antenna length, making it possible to enhance or practically eliminate one component of the overall electric field in one or more known locations \cite{blascetta20}. Swappable hot and cold spots have also been demonstrated in the nanogap of two rods by changing only the polarisation of the incident field \cite{xia20}, an example of fast, fully remote and non-mechanical control of the field around a nanoantenna. Similarly promising techniques offer ultra-fast hot spot switching between different locations near a nanostructure using Fourier limited or chirped pulses \cite{brinks13}, or by changing the direction of propagation of the incident light \cite{zhang18}. Unfortunately, hot spots only develop in very particular locations: the nanogap of a bowtie or the ends of a nanorod, for example. Tightly tied to the shape of the nanoantenna, a hot spot is not free to be moved anywhere in space. But a cold spot is. In fact, with simple and potentially ultrafast changes in the polarisation of the illuminating light, the electric fields near a nanoantenna can be designed to destructively interfere in the right places and steer a cold spot along an arbitrary path in space. This way, a cold spot’s movement can be mapped to an incident field’s polarisation signature on the Poincaré sphere.\\
\indent The components of an incident field, such as a plane wave, set up different electric fields upon interacting with the nanoantenna. By balancing the share of power and the phase difference between these components, it is possible to control some aspects of the total field surrounding a nanoantenna \cite{brixner1, brixner2} – that is, we can select the value of a component of the total field at some point in space, and find the incoming plane wave components which bring about our selection algebraically. Each plane wave component affords a handle to turn and manipulate the electric field surrounding the antenna; given enough handles, we can create a cold spot wherever we like. In fact, the sort of cold spot we want to make (having one, two or three dark components) determines the minimum number of controllable plane wave components needed incident on the antenna. Here, we show full position control of a completely dark, zero-magnitude cold spot by changing only the electric field components of two plane waves incident on two separated point scatterers. We justify this arrangement as the simplest model for consistent cold spot control, because of problematic features of the surrounding field that might appear when treating a single point scatterer. A key implication is that any arbitrary scatterer, which can be modelled as a collection of point scatterers with the discrete dipole approximation, may be used to achieve full cold spot control as long as it is large enough to escape the Rayleigh approximation (Fig. 1). We also stress that the degrees of freedom which impose the cold spot’s criteria on the total electric field could be found from sources other than the incident electric field components, including the incidence direction or even the geometry of the nanoantenna.

Lit by a plane wave $\mathbf{E}_{\textrm{a}}(\mathbf{r})\textrm{e}^{\textrm{i}\omega t}$ with propagation vector $\mathbf{k}_{\textrm{a}}$, a metal nanoparticle scatters some amount of power towards a point of interest $\mathbf{r}_{0}$. Here, the time-independent total electric field phasor, $\mathbf{E}_{\textrm{t}}(\mathbf{r}_{0})$, is the sum of the exciting and scattered field and $\mathbf{r}_{0}$ will foster a cold spot if $\mathbf{E}_{\textrm{t}}(\mathbf{r}_{0})=0$. The cold spot will remain in place even as the phase of the incident light advances over time. Then, equating the three components of $\mathbf{E}_{\textrm{t}}(\mathbf{r}_{0})$ to zero, we can build a system of three independent equations, linear with respect to the variables offered by $\mathbf{E}_{\textrm{a}}$, which – we hope – may be solved to discover the $\mathbf{k}_{\textrm{a}}$-directed plane wave which creates a cold spot at any particular location $\mathbf{r}_{0}$ points to. We aim for a non-trivial solution, meaning that the conditions which force our cold spot are held under a non-zero exciting field. This is possible if we inject at least four variables into the system, to return one or more system-wide degrees of freedom after spending the first three variables satisfying the three scalar equations implicit in $\mathbf{E}_{\textrm{t}}(\mathbf{r}_{0})=0$. These three equations are what we refer to as the linear system of equations needing solved.

\begin{figure}
\includegraphics[width=0.5\textwidth]{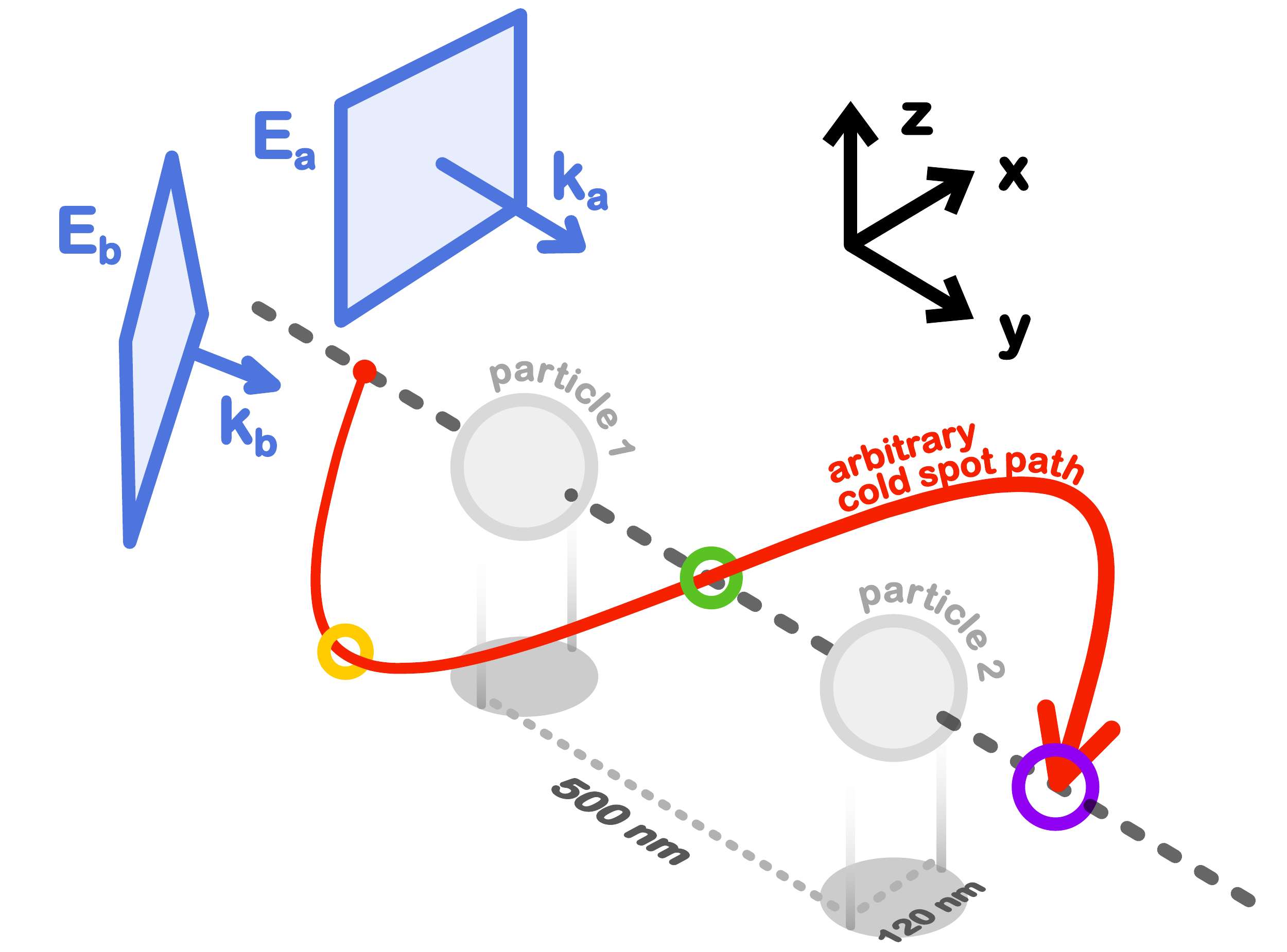}
\caption{
Illustration of the path taken by a cold spot around and in-between two silver nanoparticles, illuminated by two plane waves, $\mathbf{E}_{\textrm{a}}$ and $\mathbf{E}_{\textrm{b}}$. Coloured rings mark the three stages in the journey for which results are presented in Fig. 3.
}
\end{figure}

When, detached from our linear system of equations, we choose to make a plane electromagnetic wave $\mathbf{E}_{\textrm{a}}$ with the fixed propagation direction $\mathbf{k}_{\textrm{a}}$, we enjoy only two degrees of freedom. The transversality condition restricts the electric field vector to the plane perpendicular to $\mathbf{k}_{\textrm{a}}$, so that it is always fully expressed as the sum of two components, each being a complex amplitude, $x_{i}$, multiplied by a suitable basis vector, $\mathbf{\hat{e}}_{i}$, for $i=1, 2$.
\begin{equation}
\mathbf{E}_{\textrm{a}}(\mathbf{r})=(x_{1}\mathbf{\hat{e}}_{1}+x_{2}\mathbf{\hat{e}}_{2})\textrm{e}^{\textrm{i}\mathbf{k}_{\textrm{a}}\cdot\mathbf{r}}
\end{equation}
One degree of freedom in this context translates to one variable in our three-equation linear system $\mathbf{E}_{\textrm{t}}(\mathbf{r}_{0})=0$; a single plane wave cannot buy us a non-trivial solution. Perhaps the most obvious next step is to add another plane wave to the system, $\mathbf{E}_{\textrm{b}}(\mathbf{r})=(x_{3}\mathbf{\hat{e}}_{3}+x_{4}\mathbf{\hat{e}}_{4})\textrm{e}^{\textrm{i}\mathbf{k}_{\textrm{b}}\cdot\mathbf{r}}$. This second exciting field supplies its own two components, $x_{3}\mathbf{\hat{e}}_{3}$ and $x_{4}\mathbf{\hat{e}}_{4}$, so that the number of variables contributed to the linear system of equations (four) surpasses the number of conditions needing satisfied (three). Note that for some choices of $\mathbf{k}_{\textrm{a}}$ and $\mathbf{k}_{\textrm{b}}$, one or both basis vectors of $\mathbf{E}_{\textrm{a}}$ could be shared by $\mathbf{E}_{\textrm{b}}$, which would not dissolve any degrees of freedom if $\mathbf{k}_{\textrm{a}}\neq\mathbf{k}_{\textrm{b}}$ (which is to say that $\mathbf{E}_{\textrm{a}}$ and $\mathbf{E}_{\textrm{b}}$ are two separate plane waves). Because of the linearity of Maxwell's equations, the total field phasor evaluated at $\mathbf{r}_{0}$ can now be written as the linear sum of $x_{i}$ times the electric fields $\mathbf{E}_{i}(\mathbf{r}_{0})$ which $x_{i}$ are directly responsible for setting up (for instance, $\mathbf{E}_{3}(\mathbf{r}_{0})$ is the total field that exists at $\mathbf{r}_{0}$ when $x_{3}=1$ and all other plane wave components are zero).
\begin{equation}
\mathbf{E}_{\textrm{t}}(\mathbf{r}_{0})=x_{1}\mathbf{E}_{1}(\mathbf{r}_{0})+x_{2}\mathbf{E}_{2}(\mathbf{r}_{0})+x_{3}\mathbf{E}_{3}(\mathbf{r}_{0})+x_{4}\mathbf{E}_{4}(\mathbf{r}_{0})
\end{equation}
This equation may be written in the matrix form $\tensor{\mathbf{A}}\mathbf{x}=\mathbf{E}_{\textrm{t}}(\mathbf{r}_{0})$, where $x_{i}$ are contained in the column vector $\mathbf{x}$,
\onecolumngrid
\widetext
\begin{figure}[!htp]
\centering
\includegraphics[width=1\textwidth]{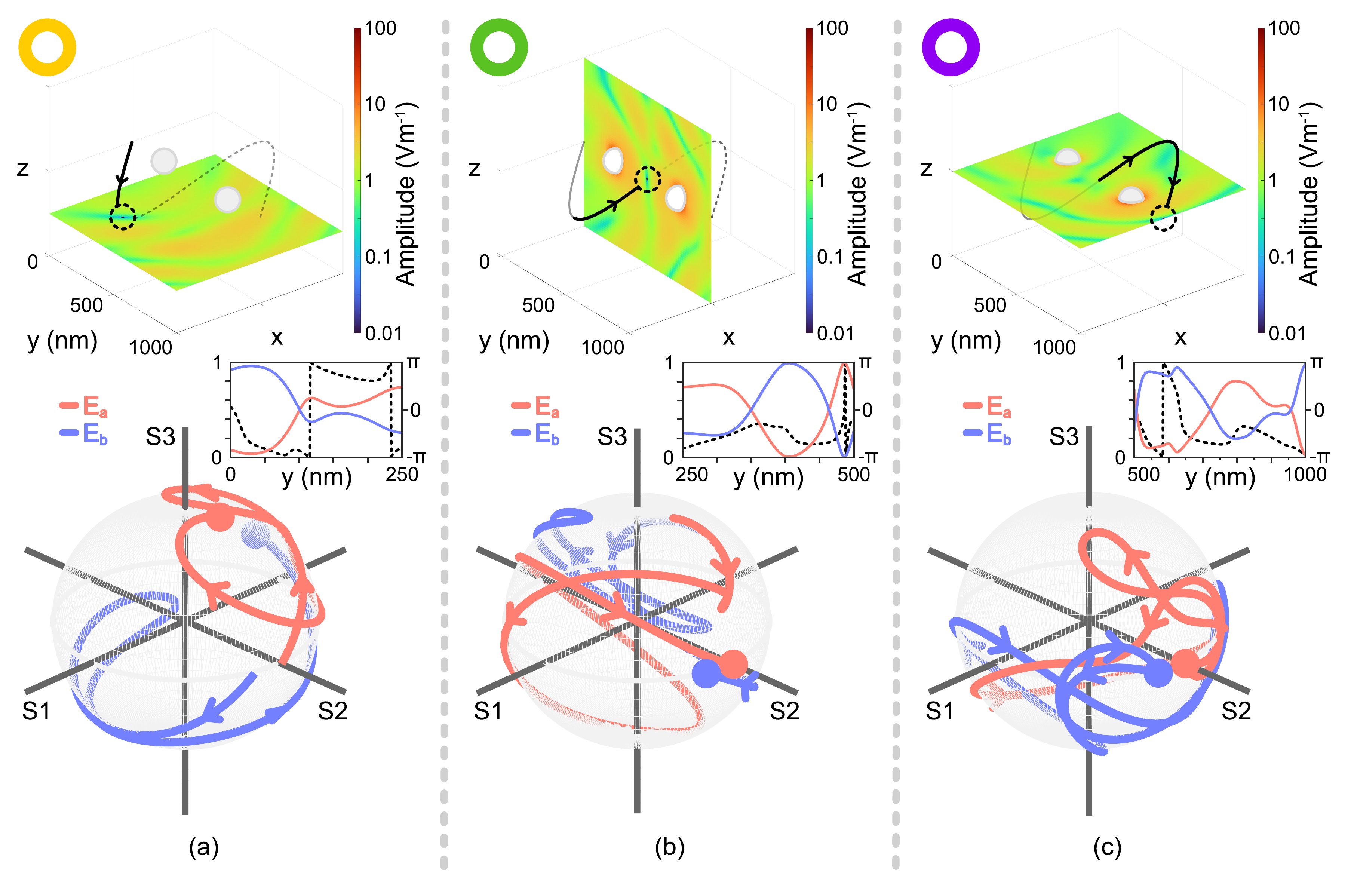}
\caption{Results obtained at the yellow (a), green (b), and purple (c) rings depicted in Fig. 2.
	Top row: Distribution of the scalar electric field amplitude surrounding the two nanoparticles of Fig. 2, plotted on an $xy$ or $yz$ plane containing the cold spot. The silver particles and cold spot’s progress along the path in Fig. 2 are drawn for visual reference.
	Bottom row: Change in the polarisation state of $\mathbf{E}_{\textrm{a}}$ (red) and $\mathbf{E}_{\textrm{b}}$ (blue), plotted on the Poincaré sphere, which moved the cold spot up to the current coloured ring from the previous (or from the beginning of the path for (a)) along the line shown in Fig. 2. The ‘head’ of the line (the red or blue circle) on the sphere is the polarisation state at the current cold spot location (shown in the field distribution), while the ‘tail’ is the polarisation state at the previous stage.
	Inset: Change in the intensity (left axis) of $\mathbf{E}_{\textrm{a}}$ (red) and $\mathbf{E}_{\textrm{b}}$ (blue) as the cold spot was moved along each of the three legs in the cold spot path, plotted versus the relevant section of the $y$ axis. The black dotted line shows the phase offset between the $\mathbf{\hat{x}}$ component of $\mathbf{E}_{\textrm{a}}$ and the $\mathbf{\hat{e}}_{\textrm{p}}$ component of $\mathbf{E}_{\textrm{b}}$ (measured by the right axis).
	}
\end{figure}
\twocolumngrid
\noindent and the vectors $\mathbf{E}_{i}(\mathbf{r}_{0})$ are the columns of the $3\times4$ coefficient matrix $\tensor{\mathbf{A}}$ evaluated at $\mathbf{r}_{0}$. In principle, it is possible to fix a cold spot at any $\mathbf{r}_{0}$ around any metal nanoparticle with the plane wave component amplitudes $x_{i}$ given by the null space of $\tensor{\mathbf{A}}$. And importantly, these component amplitudes need not be supplied by any particular number of plane waves – four TE polarised fields are an equally valid means to achieve full cold spot position control, for example. By adding even more degrees of freedom to the linear system, we can enforce extra conditions like a constant intensity for each incident field, feasible with the six variables committed by three freely polarisable plane waves. What does restrict us, however, is the particle geometry which in exceedingly simple cases can leave the mathematical solution to (2) with unwanted properties.\\
\indent Consider a very small metal particle, a point scatterer, with a point dipole at its centre induced by two plane waves travelling in different directions. The two plane waves interfere and fashion a standing wave pattern; their summed phasors produce bands of high and low intensity, skewed across the particle face by an angle arising from their propagation vectors. In this case, solutions will exist for some cold spot locations where the two plane waves’ phasors cancel out exactly at the particle centre, so that the particle never scatters power and entire cold lines or planes will be found at the nodes in the plane waves’ interference pattern. Depending on the application, we may only wish to control the position of unique and isolated, three-dimensional point cold spots, surrounded by a non-zero electric field. This problem might be avoided by modelling a larger scatterer, or by adding a second point scatterer some distance from the first which stops them both lying in a point of zero exciting intensity at the same time. It is also important to think about how cold spots might develop differently around resonant and non-resonant particles. The weak electric field scattered by a non-resonant particle barely disrupts the standing wave pattern created by the incident plane waves, such that pronounced nodes and antinodes are left to obstruct any application of this technique. Happily, a well-defined cold spot, which is not concealed in an interference pattern, can be created with the help of the stronger fields scattered by resonant particles. Drawing from the straightforward concepts in this discussion, we next demonstrate what we have now built as the simplest case for practical polarisation-controlled cold spots: a unique, super sub-wavelength cold spot is moved along an arbitrary path around two separated  point scatterers at resonance.\\
\indent The electric fields $\mathbf{E}_{i}(\mathbf{r}_{0})$ in (2) were determined for a system of two plane waves $\mathbf{E}_{\textrm{a}}$ and $\mathbf{E}_{\textrm{b}}$, with propagation vectors $\mathbf{k}_\textrm{a}=\frac{2\pi}{\lambda_{0}}\mathbf{\hat{y}}$ and $\mathbf{k}_\textrm{b}=\frac{2\pi}{\lambda_{0}}(\frac{1}{\sqrt{5}}\mathbf{\hat{x}}+\frac{\sqrt{3}}{\sqrt{5}}\mathbf{\hat{y}}-\frac{1}{\sqrt{5}}\mathbf{\hat{z}})$ respectively, incident on two identical, 120 nm diameter silver nanospheres, suspended in free space and separated along the $y$ axis by 500 nm (the spheres have co-ordinates in nm of (500, 250, 500) and (500, 750, 500)). This arrangement is illustrated in Fig. 2. The field scattered by either particle was approximated as that from a point scatterer with a polarizability tensor $\tensor{\mathbf{\boldsymbol{\upalpha}}}$. Finally, the wavelength $\lambda_{0}$ of the two incident plane waves was 354 nm where, from Palik \cite{palik}, silver meets the Fröhlich closest in free space. Derivation of the columns $\mathbf{E}_{i}(\mathbf{r}_{0})$ of the system matrix $\tensor{\mathbf{A}}$ is found for this case in the supplemental material.\\
\indent A cold spot was moved along the curving pathway pictured in Fig. 2 (an arbitrarily chosen sinusoid, advancing along $y$ and tilted with respect to the $x$ and $z$ axes), and plots of the total electric field surrounding the particles at three different stages in the cold spot’s journey (highlighted by the coloured rings in Fig. 2) are shown in Fig. 3(a – c). The field plots were drawn by solving (2) to fix a cold spot at the relevant coloured ring’s position, rounding the plane wave component amplitudes to three decimal places, summing the exciting plane waves’ and the particles’ resulting scattered fields, and computing the total electric field magnitude on a plane which reveals the desired cold spot. The null space of the coefficient matrix $\tensor{\mathbf{A}}$ was normalised so that the sum of the two plane waves’ intensities was always 1 $\textrm{V}^{2}\textrm{m}^{-2}$. The silver particles and the cold spot’s progress are drawn with an indication of the cold spot’s current location for visual reference. Plots, on a single Poincaré sphere, of the change in the plane waves’ polarisation states necessary to move the cold spot along the sinusoidal path from one coloured ring to the next accompany the field distributions. The Stokes parameters for the first plane wave are calculated with respect to $\mathbf{\hat{x}}$ and $\mathbf{\hat{z}}$ while the second plane wave’s are found using its $\mathbf{\hat{e}}_{\textrm{p}}$ and $\mathbf{\hat{e}}_{\textrm{s}}$ components, measured with respect to the positive $z$ axis. The inset traces the change in the two plane waves’ intensities and phase offset for the same sections of the cold spot’s path, which are not otherwise shown in the Poincaré spheres. An animation of the moving cold spot is provided in the supplemental material.\\
\begin{figure}
\includegraphics[width=0.5\textwidth]{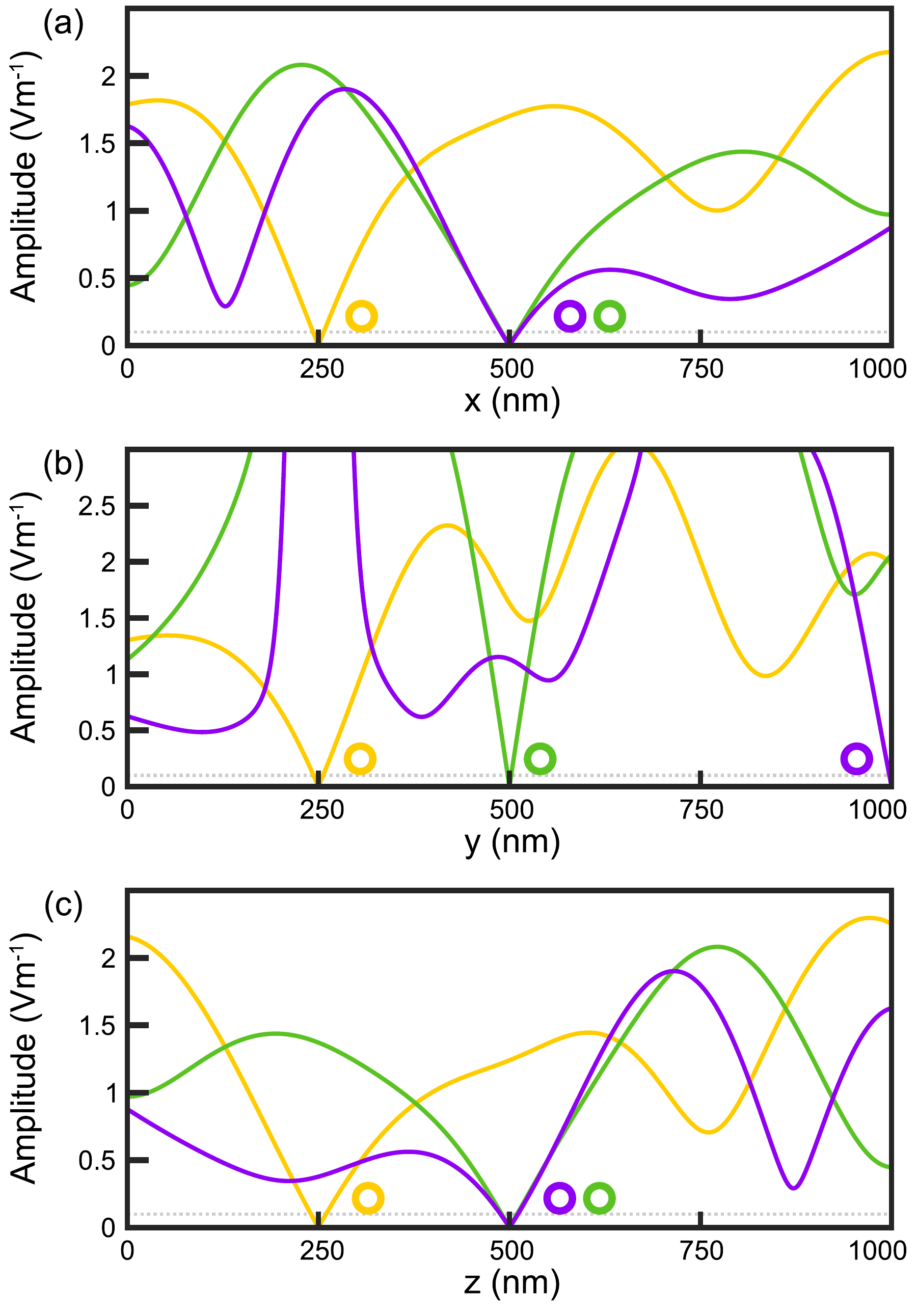}
\caption{
Cross sections of the electric field distributions of Fig. 3, showing the variation in scalar electric field amplitude along the lines parallel to $x$ (a), $y$ (b), and $z$ (c) which intersect at each of the three cold spot locations indicated by the coloured rings. In (b), the lines are clipped above 3 $\textrm{V}\textrm{m}^{-1}$ because the field diverges at 250 nm and 750 nm in the point scatterers’ centres. The grey dashed line shows the 90\% suppression level (0.1 $\textrm{V}\textrm{m}^{-1}$).
}
\label{cross sections}
\end{figure}
\indent Scrawled on the Poincaré spheres, the distinct red and blue signatures of $\mathbf{E}_{\textrm{a}}$ and $\mathbf{E}_{\textrm{b}}$ map directly to each of the four legs in the cold spot’s journey around and in-between the silver nanoparticles. Interestingly, both plane waves’ polarisations begin and end in the same place on Fig. 3(c)’s Poincaré sphere – the starting polarisation shown in Fig. 3(a). These polarisations correspond to the three points in its path that the cold spot is in line with the two particles; here, the relative intensity and phase difference between the two plane waves have the final say in which position the cold spot appears. In Fig. 4, one dimensional electric field cross sections of the domain along $x$ (a), $y$ (b), and $z$ (c) for each cold spot in the three distributions of Fig. 3 are given. In linear units, these reveal both the near complete field suppression in the cold spot and its point-like shape – even with the plane waves’ limited component precision. In fact, the 90\% suppression (of 1 $\textrm{V}\textrm{m}^{-1}$, given by the dashed line in Fig. 4) width of the cold spot is ~20 nm in the widest dip of Fig. 4, while in each cold spot centre the field was suppressed by a factor of ~1000. The electric field in the cold spot would be exactly zero had the incident plane wave components not been rounded.\\
\indent Summarising, the polarisation and relative intensities of two plane waves with fixed propagation vectors provide enough degrees of freedom to make a zero-magnitude, time-independent cold spot anywhere in the space around a nanoantenna. A simple way of realising this control is by modelling the nanoantenna as a point scatterer, and constructing and solving a linear system of equations to find the plane wave components corresponding to a set of co-ordinates for the cold spot. By adding one or more extra point scatterers, or by modelling a nanoantenna large compared to wavelength, we can avoid problematic field distributions which arise for the single point scatterer case if the two plane waves cancel out at its centre. Supplying more degrees of freedom to the system, extra conditions may be put in place in addition to those supporting the cold spot, such as a constant intensity for each incident field. Degrees of freedom may be earned from the electric field components of independent plane waves (as we have shown), or any other parameter like the incidence direction or scatterers’ relative position or geometry, when allowed to change. We have shown a way to unlock complete control of the electric field around a nanoantenna, to create unique, highly sub-wavelength and zero-magnitude cold spots anywhere in its surroundings, with simple and potentially ultrafast modulation of the incoming fields’ polarisation and intensity.

\end{document}